\documentclass{elsart}

\usepackage{epsfig}

\usepackage{amssymb}

\begin{document}

\begin{frontmatter}


% use the thanksref command within \title, \author or \address for footnotes:
% \author{Name\thanksref{label2}}
% including your email address:
% \address{Address\thanksref{email}}
% \thanks[email]{E-mail: }
\title{Measurement of direct photon emission in $K^+ \rightarrow \pi^+ 
\pi^0 \gamma$ decay using stopped positive kaons }

\thanks[email]{ Corresponding author.\\{\it E-mail address:}
 suguru@phys.wani.osaka-u.ac.jp~(Suguru SHIMIZU)}

\author[inr]{M.A.~Aliev}, 
\author[tuku1]{Y.~Asano}, 
\author[kek]{T.~Baker}, 
\author[dep]{P.~Depommier}, 
\author[has]{M.~Hasinoff}, 
\author[osa]{K.~Horie \thanksref{kate}},
\author[kek]{Y.~Igarashi},
\author[kek]{J.~Imazato}, 
\author[inr]{A.P.~Ivashkin}, 
\author[inr]{M.M.~Khabibullin}, 
\author[inr]{A.N.~Khotjantsev},
\author[inr]{Y.G.~Kudenko},
\author[inr]{A.S. Levchenko},
\author[kek]{G.Y.~Lim}, 
\author[tri]{J.A.~Macdonald}, 
\author[inr]{O.V.~Mineev}, 
\author[sas]{C.~Rangacharyulu},
\author[kek]{S.~Sawada} 
 and 
\author[osa]{S.~Shimizu \thanksref{email}}%~~\\

\collab{KEK-E470 Collaboration}
\address[inr]{
Institute for Nuclear Research,
Russian Academy of Sciences, Moscow 117312, 
      Russia }
\address[tuku1]{
 Institute of Applied Physics, University of Tsukuba, Ibaraki 305-0006, 
 Japan }
\address[kek]{
 Institute of Particle and Nuclear Studies (IPNS), High Energy Accelerator Research Organization (KEK), Ibaraki  305-0801, Japan }
\address[dep]{
Laboratoire de Physique Nucl\'eaire, Universit\'e de Montr\'{e}al,
Montr\'{e}al, Qu\'ebec, Canada H3C 3J7}
\address[has]{
 Department of Physics and Astronomy, University of British Columbia, 
 Vancouver, Canada V6T 1Z1 }
\address[osa]{Department of Physics, Osaka University, Osaka 560-0043,
 Japan }
\address[tri]{
TRIUMF, Vancouver, British Columbia, Canada V6T 2A3 }
\address[sas]{
Department of Physics, University of Saskatchewan, Saskatoon, Canada 
S7N 5E2}

\thanks[kate]{ Present address:  Institute of Particle and Nuclear Studies (IPNS), High Energy Accelerator Research Organization (KEK), Ibaraki  305-0801, Japan }

\begin{abstract}
The radiative decay $K^+ \rightarrow \pi^+ \pi^0 \gamma$ 
($K_{\pi 2 \gamma}$) has been measured with stopped positive kaons.
%in order to study direct photon emission process. 
A $K_{\pi 2 \gamma}$ sample containing 4k events was analyzed, and the 
$K_{\pi 2 \gamma}$ branching ratio of the direct photon emission process was 
determined to be $[6.1\pm2.5({\rm stat})\pm1.9({\rm syst})]\times 10^{-6}$.
No interference pattern with internal bremsstrahlung was observed.
\end{abstract}

%\begin{keyword}
%\sep $K_{e3}$
%\PACS 13.20.Eb
%\end{keyword}

\end{frontmatter}
\section{Introduction}
The chiral anomaly~\cite{adl69}, which is a basic feature of quantum field 
theory, has been the subject of extensive theoretical investigations. There 
are a number of calculations for decays of the pseudoscalar mesons -- $\pi$, 
$\eta$, and $K$~\cite{don89}; however few
experimental studies of this anomaly have been performed so far. With the 
increased interest in chiral perturbation theory (ChPT)~\cite{bur96}, 
the chiral anomaly has attracted renewed
attention. In particular, non-leptonic kaon decays are well-suited
for the study of ChPT. It is known, however, that the genuine  
manifestation of the chiral anomaly in non-leptonic decays is restricted 
to the radiative
decay of $K^+ \rightarrow \pi^+ \pi^0 \gamma$ ($K_{\pi 2 \gamma}$) 
for the charged kaon~\cite{eck94}.

The $K^+ \rightarrow \pi^+ \pi^0$ ($K_{\pi 2}$) decay is hindered because 
it violates the 
$\Delta I$=1/2 rule. As a consequence, the internal 
bremsstrahlung (IB) contribution to  $K_{\pi 2 \gamma}$
decay is suppressed, although it is still dominant. This feature, 
in turn, could enhance the direct photon emission
(DE) which is sensitive to meson structure. Here, the chiral
anomaly appears as the magnetic component of DE.  In terms of
ChPT, it enters the magnetic amplitude at $O(p^4$), while IB arises
from $O(p^2)$~\cite{eck94}.  Also, the electric amplitude of DE may arise 
from $O(p^4$), which one can identify 
through the interference pattern (INT) with IB. 
Thus the $K_{\pi2 \gamma}$ decay channel is one of the most interesting
and important channels in determining the low energy structure of QCD.

Three in-flight-$K^+$ experiments~\cite{abr72,smi76,bol87} and one 
stopped-$K^+$ experiment~\cite{adl00} have reported on the DE branching 
ratio in 
the $\pi^+$ kinetic energy region of $55<T_{\pi^+}<90$ MeV. These experiments 
restricted the $T_{\pi^+}$ region to avoid backgrounds from 
$K^+ \rightarrow \pi^+ \pi^0 \pi^0$ ($K_{\pi 3}$) decay, although 
the DE component is significantly enhanced with decreasing $\pi^+$ 
energy. The weighted average of the 
in-flight $K^+$ experiments is $Br(K_{\pi 2 \gamma},{\rm DE})=(18\pm4)
\times 10^{-6}$~\cite{par00} which is much larger than the 
result from the stopped $K^+$ experiment $Br(K_{\pi 2 \gamma},{\rm DE})=
(4.7\pm0.8\pm0.3)\times 10^{-6}$~\cite{adl00}. 
Therefore, further
experimental studies are necessary to settle this discrepancy
of the DE branching ratio. It is also desirable to check the existence of 
the INT component which has not yet been observed.

In this letter, we present a new measurement of direct photon emission
in $K_{\pi 2 \gamma}$ decay. The experiment used a stopped $K^+$ beam
in conjunction with a superconducting toroidal 
spectrometer with a 12-sector iron-core~\cite{ima:90}. We measured the 
$K_{\pi 2 \gamma}$ events by extending 
the $T_{\pi^+}$ region ($T_{\pi^+}>$35 MeV) below the $K_{\pi 3}$ upper 
threshold ($T_{\pi^+}^{\rm max}=$55 MeV).

\section{Experiment}
\begin{figure}
\begin{center}
\epsfig{file=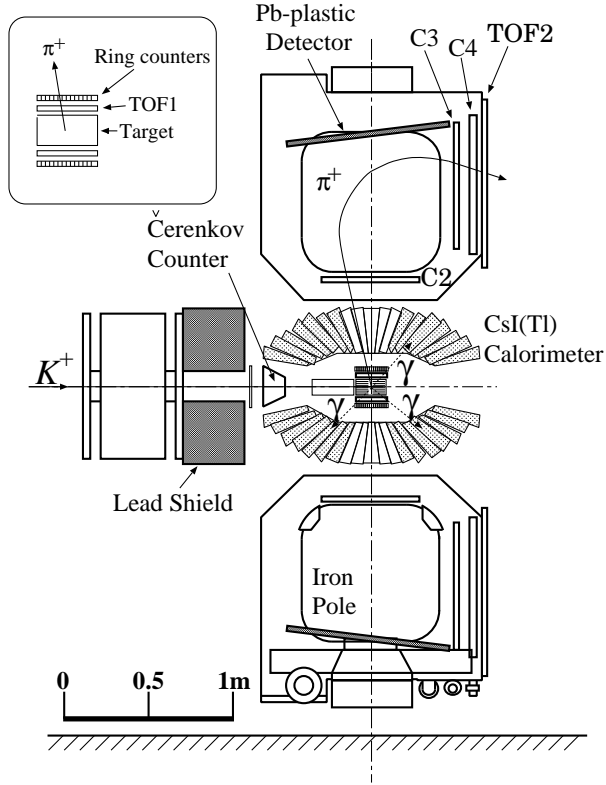,height=10.5cm,width=8.3cm}
\caption{ 
Cross sectional side view of the E470 setup. Assembly detail of the active
 target, TOF1, and ring counters is shown in the inset.
}
\label{fg:side}
\end{center}
\end{figure}

The experiment was performed at the KEK 12 GeV proton synchrotron. 
The experimental apparatus was based on the E246 experiment 
searching for $T$-violating transverse muon polarization in 
$K^+\rightarrow\pi^0 \mu^+ \nu$ ($K_{\mu 3}$) decay~\cite{main:99}. 
Besides the $T$-violation search, spectroscopic studies for 
$K^+\rightarrow\pi^0 e^+ \nu$ ($K_{e 3}$), $K_{\mu 3}$, and $K_{\pi 3}$
decays have been carried out using the same detector 
system~\cite{shi00,hor01,ysh00}. A schematic cross 
sectional view of the detector is shown in Fig.~\ref{fg:side}. 
Because of the rotational symmetry of the 
12 identical gaps in the spectrometer and the large directional 
acceptance of the photon detector~\cite{nimcsi}, 
distortions of the $K_{\pi 2 \gamma}$
spectra due to instrumental misalignment, if any, were drastically reduced.

A separated 660 MeV/$c$ $K^+$ beam ($\pi^+/K^+ \sim 7$) was used. Each
$K^+$ was discriminated from pion background by a ${\rm \check{C}}$erenkov 
counter with an 
efficiency of more than 99\%. The typical $K^+$ beam intensity was
$1.5\times10^5$ over a 2 s spill duration with a 4 s repetition rate. The
kaons were slowed down by a degrader and stopped in an active target 
located at the center of the detector system.
The $K_{\pi 2 \gamma}$ events were identified by analyzing the $\pi^+$ 
momentum with the spectrometer and detecting three photons in the 
CsI(Tl) calorimeter. Charged particles from the target were tracked
and momentum-analyzed using multi-wire proportional chambers  at
the entrance (C2) and exit (C3 and C4) of the magnet gap, as well as 
an array of ring counters~\cite{ring:97} surrounding the active target
system. The $\pi^+$s were discriminated from $e^+$ 
and $\mu^+$ by calculating the mass squared ($M_{\rm TOF}^2$) of the 
charged particles from the time-of-flight between TOF1 and TOF2 counters. 
Counter TOF1 surrounded 
the active target and counter TOF2 was located at the exit of the spectrometer.
The measurement was carried out for the central magnetic field strength
B=0.65 T which was chosen to match the spectrometer acceptance to the 
$\pi^+$ momentum distribution due to the DE process.

The photon detector, an assembly of 768 CsI(Tl) crystals, covered 75\%
of the total solid angle. There were 12 holes for outgoing charged 
particles to enter the spectrometer and 2 holes for beam entrance and 
exit, as shown in Fig.~\ref{fg:side}. 
The photon energy and hit position were obtained, respectively, 
by summing the energy deposits and taking the energy-weighted centroid
of the crystals sharing a shower. Timing information from each module was 
used to identify a
photon cluster and to suppress accidental backgrounds due to beam
particles. A Pb-plastic sandwich detector with 2.6 radiation length was set
at the outer radius of the magnet pole to monitor photons passing through 
the photon detector holes, as shown in Fig.~\ref{fg:side}.

\section{$K_{\pi2 \gamma}$ event selection}
\begin{figure}
\begin{center}
\epsfig{file=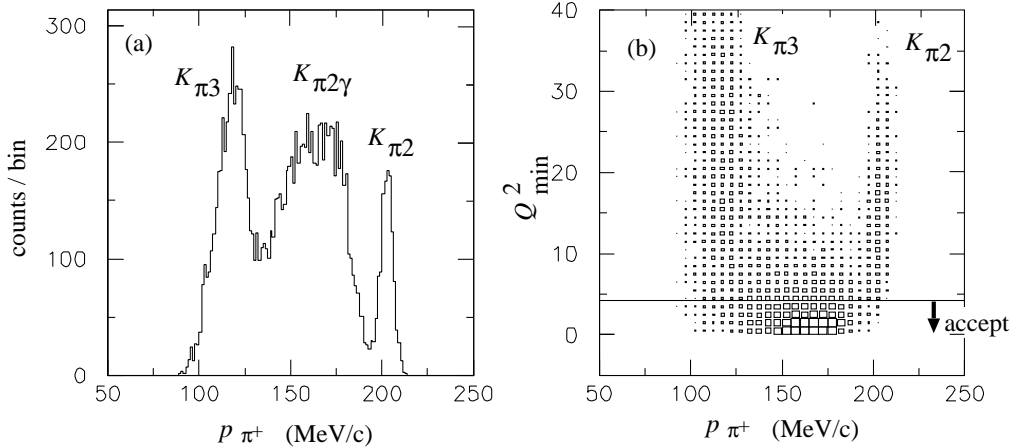,height=6.0cm,width=13.5cm}
\caption{(a) $p_{\pi^+}$ distribution and (b) correlation 
plot of $p_{\pi^+}$ and $Q^2_{\rm min}$ without 
$p_{\pi^+}$ and $Q^2_{\rm min}$ cuts. To demonstrate 
background rejection performance by the $Q_{\rm min}$ cut, 
the $p_{\pi^+}$ cut is not applied here.
}
\label{fg:eve}
\end{center}
\end{figure}

$K_{\pi2 \gamma}$ decays at rest were obtained by the following 
procedure which was carefully studied in a Monte Carlo simulation
(see below). 
The $K^+$ decay time, defined as the $\pi^+$ signal at the 
TOF1 counter, was required to be more than 1.4 ns later than the $K^+$
arrival time measured by the ${\rm \check{C}}$erenkov counter. This reduced
the fraction of $K^+$ decays in-flight contamination to the level of 
$10^{-3}$. 
Events with $\pi^+$ decays in-flight and scattering of charged particles
from the magnet 
pole faces were eliminated by requiring
the hit position in ring counters to be consistent with
the charged particle track.
The selection of $\pi^+$, by $15000<M_{\rm TOF}^2<30000$ MeV$^2/c^4$, 
rejected $e^+$s and $\mu^+$s. 
To remove $K_{\pi 2}$ events, the reconstructed $\pi^+$ momentum $p_{\pi^+}$
(Fig.~\ref{fg:eve}(a)), corrected for the 
energy loss in the target, was required to be $p_{\pi^+}<$175 
MeV/$c$. Events with three clusters in the CsI(Tl) calorimeter 
were selected by assigning
two photons from $\pi^0 \rightarrow \gamma_1+\gamma_2$ and a radiated photon
($\gamma_3$), while events with other cluster number were rejected. 
The opening angle
between the two photons was required to be larger than 37$^\circ$
to eliminate photon conversion backgrounds in the active target system. 
Also, the photon conversion events could be removed by selecting only 
events in 
which there was one charged particle hit in TOF1 as well as in ring 
counters.
The total $K^+$ energy $M_{K^+}=E_{\pi^+}+\sum_{i=1}^{3}E_{\gamma_{i}}$ 
and momentum vector $\mbox{\boldmath $p$}_{K^+}=
\mbox{\boldmath $p$}_{\pi^+}
+\sum_{i=1}^{3}\mbox{\boldmath $p$}_{\gamma_{i}}$ 
were used to reject background contaminations. The windows of 
$420<M_{K^+}<500$ MeV/$c^2$ and  
$|(\mbox{\boldmath $p$}_{K^+})_{x,y,z}|<$50 MeV/$c$ were imposed.

The final step in reconstructing $K_{\pi2 \gamma}$ was to determine
which photons were a pair from a $\pi^0$, because there are
three possible combinations to form the $\pi^0$. The IB
events with incorrect pairing mimic the DE decay
and might introduce a systematic error. In order to find the 
correct pairing, a quantity $Q^2$ defined as,
\begin{eqnarray}
Q^2&=&(M_{\gamma_1 \gamma_2}-\widetilde{M}_{\pi^0})^2/\sigma^2_{M_{\pi^0}}+ (E_{\pi^0}+E_{\gamma_3}+E_{\pi^+} -\widetilde{M}_{K^+})^2/\sigma^2_{M_{K^+}}\nonumber \\ 
&+&({\rm cos}\theta_{\pi^+ \gamma_3}^{\rm CALC}-{\rm cos}\theta_{\pi^+ \gamma_3}^{\rm MEAS}- \delta_1)^2/\sigma^2_{\pi^+ \gamma_3} \nonumber \\
&+& 
({\rm cos}\theta_{\pi^+ \pi^0}^{\rm CALC}-{\rm cos}\theta_{\pi^+ \pi^0}^{\rm MEAS}- \delta_2)^2/\sigma^2_{\pi^+ \pi^0}, \label{eqq2}
\end{eqnarray}
was calculated for the three combinations, and the pair with the 
minimum $Q^2$, $Q^2_{\rm min}$, was adopted as the $\pi^0$ pair. Here,
$M_{\gamma_1 \gamma_2}$ is the invariant mass of the selected pair and 
$E_{\pi^0}$ is the $\pi^0$ energy obtained as,
\begin{eqnarray}
E_{\pi^0}=\sqrt
{ E_{\gamma_1}^2+2E_{\gamma_1}E_{\gamma_2}{\rm cos}\theta_{\gamma_1 \gamma_2}+E_{\gamma_2}^2+m_{\pi^0}^2},
\end{eqnarray}
where $\theta_{\gamma_1 \gamma_2}$ is the opening angle of the selected pair
and $m_{\pi^0}$ is the rest mass of $\pi^0$. 
The angles $\theta_{\pi^+ \gamma_3}$ 
and $\theta_{\pi^+ \pi^0}$ are that between $\pi^+$  and $\gamma_3$
and that between $\pi^+$ and $\pi^0$, respectively. 
The superscripts MEAS and CALC stand 
for the measured angles and the calculated angles using the $K_{\pi2
\gamma}$ kinematics as,
\begin{eqnarray}
{\rm cos }\theta_{\pi^+ \gamma_3}^{\rm CALC}&=&\frac{ m_{K^+}^2+m_{\pi^+}^2-m_{\pi^0}^2-2 m_{K^+}E_{\pi^+}-2 m_{K^+}E_{\gamma_3}+2E_{\gamma_3}E_{\pi^+}}{ 2|\mbox{\boldmath $p$}_{\pi^+}||\mbox{\boldmath $p$}_{\gamma_3}|},\\
{\rm cos }\theta_{\pi^+ \pi^0}^{\rm CALC}&=&\frac{ m_{K^+}^2+m_{\pi^+}^2+m_{\pi^0}^2-2 m_{K^+}E_{\pi^+}-2 m_{K^+}E_{\pi^0}+2E_{\pi^0}E_{\pi^+}}{ 2|\mbox{\boldmath $p$}_{\pi^+}||\mbox{\boldmath $p$}_{\pi^0}|},
\end{eqnarray}
where $m$, $E$, and $\mbox{\boldmath $p$}$ denote the rest mass, 
energy, and momentum
vector, respectively, of the particle shown in the subscript. 
The $\sigma$ and offset values for each term were 
$\sigma_{M_{\pi^0}}$= 11.4 MeV/$c^2$, $\sigma_{M_{K^+}}$=10.5 MeV/$c^2$, 
$\sigma_{\pi^+ \gamma_3}$=0.48, $\sigma_{\pi^+ \pi^0}$=0.06,
$\widetilde{M}_{\pi^0}= 116.0$ MeV/$c^2$, $\widetilde{M}_{K}= 471.6$ MeV/$c^2$,
$\delta_1$=0.35, and $\delta_2$=0.06. The choice of Eq.~(\ref{eqq2}) and
these parameters was determined to get the highest probability for the 
correct pairing by using a simulation with the IB component.
The correct pairing probability 
was estimated to be 85\% for IB, 79\% for DE, and
83\% for INT. Since $K_{\pi3}$ and
$K_{\pi 2}$ decays do not satisfy the $K_{\pi 2 \gamma}$ kinematics, the
$Q^2_{\rm  min}$ values of the $K_{\pi3}$ and $K_{\pi 2}$ events are 
essentially larger, as shown in Fig.~\ref{fg:eve}(b). Therefore, the cut of
$Q^2_{\rm min}<4.3$
drastically reduced these contaminations.

\begin{figure}
\begin{center}
\epsfig{file=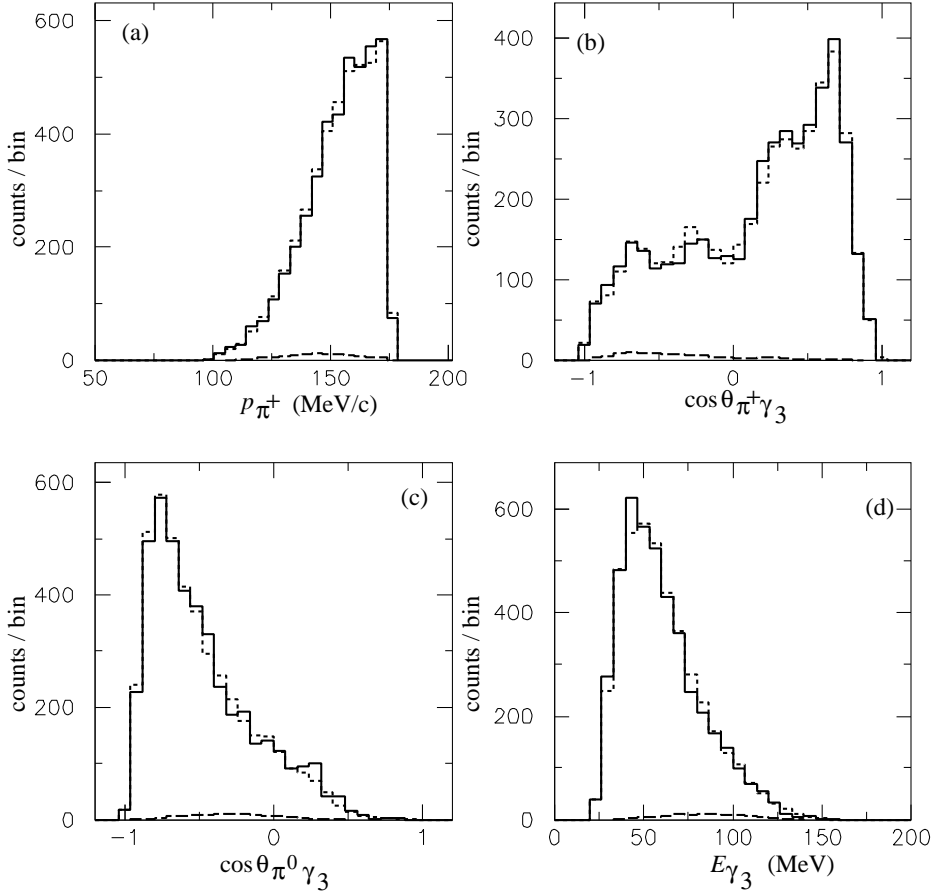,height=12.0cm,width=12.4cm}
\caption{ $K_{\pi 2 \gamma}$ spectra: (a) $\pi^+$ momentum, (b) opening 
angle between $\pi^+$ and $\gamma_3$, (c) opening angle 
between $\pi^0$ and $\gamma_3$, 
(d) $\gamma_3$ energy. The solid lines are the experimental data.
The dotted and dashed lines show the fitted spectra
and the sole DE components, respectively,
under the assumption of no INT component. The bump structure in (b) is
due to the CsI(Tl) assembly structure with 12 holes.
}
\label{fg:res}
\end{center}
\end{figure}

The number of good $K_{\pi 2 \gamma}$ events is
4434 after the above standard selection conditions. The background fraction of the 
$K_{\pi 3}$ decay was estimated to be 1.2\% from the simulation. The 
solid lines in Fig.~\ref{fg:res} show the experimental 
$K_{\pi 2 \gamma}$ spectra  after subtracting the 
backgrounds: (a) $\pi^+$ momentum, (b) opening angle between
$\pi^+$ and $\gamma_3$, (c) opening angle 
between $\pi^0$ and $\gamma_3$, and (d) $\gamma_3$ energy. 
These are compared with the calculated ones (see below).

\section{Monte Carlo simulation}
In order to obtain the detector response function, a
Monte Carlo simulation was carried out for both the $\pi^+$ measurement by the
spectrometer and the photon measurement by the CsI(Tl)
calorimeter using a GEANT-based Monte Carlo code.  
In this simulation, $K_{\pi 2 \gamma}$ events due to the IB, 
DE, and INT processes were generated with the corresponding matrix 
elements~\cite{daf92} and analyzed in the same manner as the experimental 
data.
The accuracy of the simulation was carefully checked by comparing the
$K_{\pi 2}$ spectrum with the experimental one. 
The $K_{\pi 3}$ decay was also used to check the
reproducibility of the experimental conditions in the simulation.

%The detector acceptance was obtained, respectively, to be $\Omega({\rm IB})=
%0.24\times 10^{-3}$, $\Omega({\rm DE})=0.63
%\times 10^{-3}$, and $\Omega({\rm INT})=0.61\times 10^{-3}$.
The detector acceptance ratios were obtained to be 
$\Omega({\rm DE})/\Omega({\rm IB})=2.63$ and $\Omega({\rm INT})
/\Omega({\rm IB})=2.56$ which were used 
to determine relative decay
strength of the DE and INT components normalized to the IB component.
%for the determination of the DE and 
%INT branching ratios.
%The $\Omega({\rm DE})/\Omega({\rm IB})$ and $\Omega({\rm INT})
%/\Omega({\rm IB})$ ratios were used for the determination of the DE and 
%INT branching ratios.
Here it should be noted that events with incorrect pairing were 
included in the $K_{\pi 2 \gamma}$ sample, and this effect
was also treated in the simulation.
The pairing probability could be controlled by requiring extra cuts in 
addition to the standard cuts at the cost of good
$K_{\pi 2 \gamma}$ events. 
\newcounter{aa}
Two cut conditions, \setcounter{aa}{1}(\Roman{aa})
$Q^2_{\rm sec}-Q^2_{\rm min}>1$ and \setcounter{aa}{2}(\Roman{aa})
 $\cos \theta_{\gamma_1
\gamma_2} > 0 $ or $\cos \theta_{\pi^0 \gamma_3} < 0 
$, were individually or simultaneously imposed, where 
$Q^2_{\rm sec}$ is the second smallest $Q^2$ value among the three
combinations. The pairing probability and good event acceptance under
these conditions are summarized 
in Table \ref{tb:peff}. It is possible to
increase the probability up to 95\% (IB) and 85\% (DE), although 
23\% of the $K_{\pi 2\gamma}$ events are rejected. These cuts
were used for the estimation of systematic errors.

\begin{table}
\begin{center}
\caption{Summary of the detector acceptance $\Omega$ and correct 
pairing probability $P$ for IB and DE.
}
\begin{tabular}{lcccc}
\hline 
Condition& $\Omega$(IB) 
& $\Omega$(DE) 
& $P$(IB) & $P$(DE) \\
&($10^{-3}$) & ($ 10^{-3}$)  & (\%)& (\%)\\
\hline
standard  & 0.24& 0.63& 85& 79\\
standard$\otimes$\setcounter{aa}{1}(\Roman{aa}) & 0.20& 0.53& 91& 84\\
standard$\otimes$\setcounter{aa}{2}(\Roman{aa}) & 0.22& 0.53& 92& 80\\
standard$\otimes$\setcounter{aa}{1}(\Roman{aa})$\otimes$\setcounter{aa}{2}(\Roman{aa}) & 0.18& 0.46& 95& 85\\
\hline 
\end{tabular}
\label{tb:peff}
\end{center}
\end{table}

\begin{figure}
\begin{center}
\epsfig{file=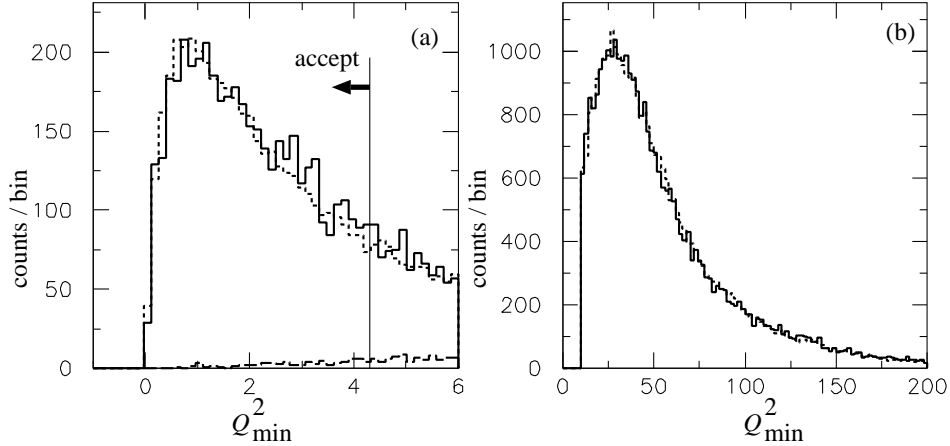,height=6.0cm,width=12.6cm}
\caption{ $Q^2_{\rm min}$ distributions of (a) $K_{\pi 2 \gamma}$
and (b) $K_{\pi3}$ events for the experimental data
(solid line) and the simulation (dotted line). Dashed line in (a)
shows the simulated $K_{\pi 3}$ background
survived after the $K_{\pi 2 \gamma}$ selection.
}
\label{fg:qdis}
\end{center}
\end{figure}

The $K_{\pi 3}$ background was also estimated 
by the simulation and subtracted from the experimental data. 
Since the present spectrometer acceptance overlapped the $K_{\pi 3}$ region, 
the estimation of
the $K_{\pi 3}$ fraction was important. The obvious $K_{\pi 3}$
events were identified as 3 photon clusters with one escaping photon
using cuts of $T_{\pi^+}<$ 55 MeV ($p_{\pi^+}<135$ 
MeV/$c$), $350<M_{K^+}<440$ MeV/$c^2$, 
$15000<M_{\rm TOF}^2<35000$ MeV$^2/c^4$,
and $Q^2_{\rm min}>10$. These events 
were compared with the simulation to confirm the accuracy of the
simulation and to check the correctness of the background estimation. 
Fig.~\ref{fg:qdis}(a),(b) show the $Q^2_{\rm min}$ distributions of 
the $K_{\pi 2 \gamma}$
and $K_{\pi3}$ events, respectively, for the experimental data
(solid line) and the simulation (dotted line). They are in excellent 
agreement, supporting the correct estimation
of the $K_{\pi 3}$ fraction in the $K_{\pi 2 \gamma}$ sample.
The dashed line in Fig.~\ref{fg:qdis}(a) shows the simulated 
$K_{\pi 3}$ background which survived after the $K_{\pi 2 \gamma}$ selection.

\section{Results}
\begin{figure}
\begin{center}
\epsfig{file=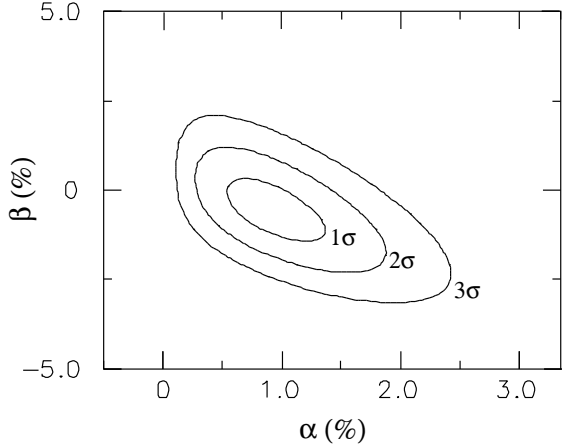,height=6cm,width=7.5cm}
\caption{ 
$\chi^2$ contour plot in the ($\alpha$, $\beta$) space. Contour levels 
of 1-, 2-, and 3-$\sigma$ 
constraints are drawn. 
}
\label{fg:con}
\end{center}
\end{figure}

\begin{figure}
\begin{center}
\epsfig{file=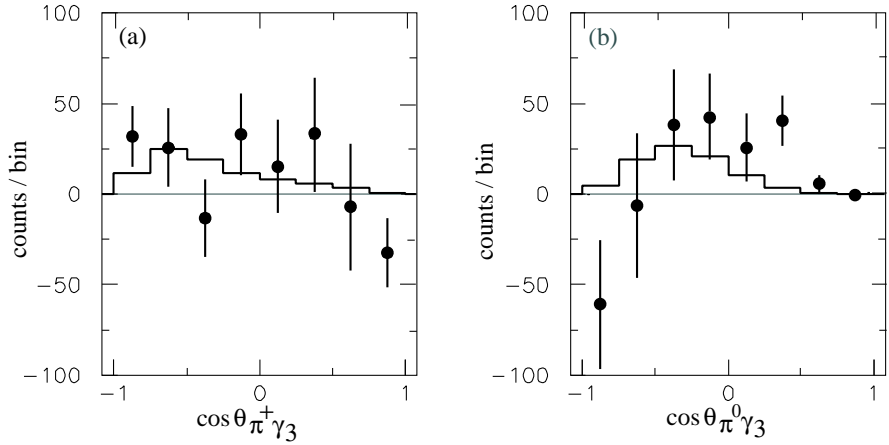,height=6cm,width=11.8cm}
\caption{ 
(a) cos$\theta_{\pi^+ \gamma_3}$ and (b) cos$\theta_{\pi^0 \gamma_3}$ 
spectra of the DE component obtained by subtracting 
the fit IB component from the experimental data (dots) and the fitted 
DE component (histogram).
}
\label{fg:sub}
\end{center}
\end{figure}

The DE branching ratio was determined using the three observables,
cos$\theta_{\pi^+ \gamma_3}$, 
cos$\theta_{\pi^0 \gamma_3}$, and $E_{\gamma_3}$  which have 
characteristic spectra corresponding to the decay processes.
The experimental distribution  
$\rho({\rm cos}\theta_{\pi^+ \gamma_3 }, {\rm cos}\theta_{\pi^0 \gamma_3}, 
E_{\gamma_3} )$ of $K_{\pi2 \gamma}$ data from the standard
cut conditions was fitted to
a simulation spectrum of IB[1+$\alpha$(DE/IB)+$\beta$(INT/IB)] with 
$\alpha$ and $\beta$
being free parameters. The DE and INT components were obtained
to be $\alpha$=(0.92$^{+0.44}_{-0.38}$)\% 
and $\beta$=($-0.58^{+0.91}_{-0.83}$)\%,
respectively. The reduced $\chi^2$ is 1.27.
Fig.~\ref{fg:con} shows the $\chi^2$ contour plot in
the ($\alpha$, $\beta$) space. $\beta$ is consistent with zero  and we  
conclude that there is no significant INT component observed within 
the experimental accuracy.
By assuming $\beta=0$ in the fitting, the DE component was 
determined to be $\alpha$=(0.76$\pm$0.31)\% with a reduced 
$\chi^2$=1.28. The DE branching ratio
was derived to be $Br(K_{\pi 2 \gamma}, {\rm DE})$=(6.1$\pm$2.5)
$\times 10^{-6}$ by normalizing it
to the theoretical value of the IB branching ratio~\cite{daf92}. 
The dotted and dashed lines in Fig.~\ref{fg:res} show the fitted spectra 
and the sole DE components, respectively,
under the assumption of $\beta=$0.  It can be seen that 26\% of DE comes
from the region $T_{\pi^+}<55$ MeV. 
Fig.~\ref{fg:sub}(a),(b) show the cos$\theta_{\pi^+ \gamma_3}$ and  
cos$\theta_{\pi^0 \gamma_3}$ spectra, respectively, of the DE 
component obtained by 
subtracting the fitted IB component from the experimental data (dots) and 
the fitted DE component (histogram).

The major systematic errors in the DE measurement come mainly from imperfect
reproducibility of the experimental conditions in the simulation as follows. 
The $\gamma$ mispairing effect was studied by varying the pairing 
probability. In addition to the standard $K_{\pi 2\gamma}$ selection conditions, 
extra cuts given in Table.\ref{tb:peff}
were imposed. The DE branching ratio was found to be consistent 
within $\Delta \alpha<$0.17\% which was regarded as the upper 
bound of this systematic error.
The uncertainty due to the calibration error of the CsI(Tl) calorimeter was
evaluated 
%in the change of DE branching ratio 
by introducing the maximum conceivable errors in the simulation, 
and found to be $\Delta \alpha=0.13$\%. The $K_{\pi 3}$ 
fraction in the $K_{\pi 2 \gamma}$ data depends slightly on the $K_{\pi 3}$
decay form factors, however this effect is negligibly small. 
The systematic error
due to the ambiguity of the $K_{\pi 3}$ fraction was obtained as
$\Delta \alpha=0.03$\% by rejecting events with 
$T_{\pi}<$55 MeV. The contribution from the accidental background 
was estimated to be $\Delta \alpha=0.08$\% by changing its fraction
varying the TDC gate widths. The uncertainty due to the spectrometer
field ambiguity was obtained to be $\Delta \alpha=0.02$\%.
As shown in Table \ref{tb:sys}, the total 
systematic error was estimated to be $\Delta \alpha_{\rm total}<$0.23\% by 
adding all the item in quadrature. However, this upper bound was regarded
here as one standard deviation of the systematic error.

\begin{table}
\caption{Summary of estimated systematic uncertainty. All items are
 added in quadrature to get the total.}
\begin{center}
\begin{tabular}{lc}
\hline 
 Error source& Uncertainty of $\alpha$ (\%) \\ 
\hline 
 $\gamma$ mispairing & $<$0.17\\
 $\gamma$ calibration & 0.13\\
 $K_{\pi 3}$ background& 0.03\\
 Accidental background & 0.08\\
 Spectrometer field ambiguity& 0.02\\
\hline
 Total systematic error &$<$0.23 \\
\hline 
\end{tabular}
\label{tb:sys}
\end{center}
\end{table}

\section{Conclusion}
We have performed a new measurement of the $K^+ \rightarrow \pi^+ \pi^0
\gamma$ decay by extending the $T_{\pi^+}$ lower bound down to 35 MeV.
The detector response and acceptance functions were evaluated
by a Monte Carlo simulation based on GEANT. The reproducibility of the 
experimental conditions in the simulation was carefully checked using 
$K_{\pi 2}$ and $K_{\pi 3}$ decays. The $K_{\pi 3}$ background 
contamination in the $K_{\pi 2 \gamma}$ data sample was also estimated by 
the simulation.  

Comparing the experimental data with the simulation, the DE component 
was obtained to be 
\begin{eqnarray}
\alpha=[0.76\pm 0.31({\rm stat}) \pm 0.23({\rm syst})]\%, \nonumber
\end{eqnarray}
under the assumption of no INT component. The DE branching ratio was then 
determined to be 
\begin{eqnarray}
 Br({\rm DE}, {\rm total})=[6.1 \pm 2.5({\rm stat}) \pm 1.9({\rm syst})]\times 10^{-6}, \nonumber 
\end{eqnarray}
by normalizing it to the theoretical value of the IB branching ratio~\cite{daf92}.
In order to compare this result with the previous ones, the partial 
branching ratio in the region of $55<T_{\pi^+}<90$
MeV was also calculated as  
\begin{eqnarray}
  Br({\rm DE},{\rm partial})=[3.2 \pm 1.3({\rm stat}) \pm 1.0({\rm syst})]\times 10^{-6}, \nonumber
\end{eqnarray}
which is consistent with the previous stopped experiment~\cite{adl00}.
%and inconsistent with the previous in-flight experiments. 
We did not  observe any significant effect due to the INT component, 
indicating the pure magnetic nature of the direct photon emission process
in $K_{\pi2\gamma}$ decay.

\begin{ack}
This work has been supported in Japan by a Grant-in-Aid from the Ministry 
of Education, Culture, Sports, Science and Technology, and by JSPS; in 
Russia by
the Ministry of Science and Technology, and by the Russian Foundation
for Basic Research; in Canada by NSERC and IPP, and by the TRIUMF
infrastructure support provided under its NRC contribution.
The authors gratefully acknowledge the excellent support
received from the KEK staff.
\end{ack}

\end{document}